# Mid-Infrared Plasmonic Platform based on Heavily Doped Epitaxial Ge-on-Si: Retrieving the Optical Constants of Thin Ge Epilayers

Leonetta Baldassarre[1], Eugenio Calandrini[2], Antonio Samarelli[3], Kevin Gallacher[3], Douglas J. Paul[3],
Jacopo Frigerio[4], Giovanni Isella[4], Emilie Sakat[5] Marco Finazzi[5],
Paolo Biagioni[5] and Michele Ortolani[2]

[1]Center for Life NanoScience@Sapienza, Istituto Italiano di Tecnologia, Rome, Italy
[2] Physics Department, Sapienza University of Rome, Italy
[3]School of Engineering, University of Glasgow, U.K.
[4]L-NESS, Politecnico di Milano, Como, Italy
[5]Physics Department, Politecnico di Milano, Italy

*Abstract*— The n-type Ge-on-Si epitaxial material platform enables a novel paradigm for plasmonics in the mid-infrared, prompting the future development of lab-on-a-chip and subwavelength vibrational spectroscopic sensors. In order to exploit this material, through proper electrodynamic design, it is mandatory to retrieve the dielectric constants of the thin Ge epilayers with high precision due to the difference from bulk Ge crystals. Here we discuss the procedure we have employed to extract the real and imaginary part of the dielectric constants from normal incidence reflectance measurements, by combining the standard multilayer fitting procedure based on the Drude model with Kramers-Kronig transformations of absolute reflectance data in the zero-transmission range of the thin film.

## I. Introduction And Background

**M**ID-INFRARED (mid-IR) sensors based on resonant absorption through specific vibrational excitations in molecules may become a fundamental tool for biology, chemistry, medicine and safety & security, after the recent development of tunable mid-IR quantum cascade lasers. Plasmonics is the most promising approach to achieve deep sub-wavelength concentration of optical fields towards single-molecule sensing. At variance with radio-frequency antennas treated in the perfect-electric-conductor limit, plasmonic antennas produce intense localized energy spots at electromagnetic frequencies close to the plasma frequency $\omega_p$ of the conductor they are made of [1].

So far, plasmonic applications have been based on metals such as gold, silver, or aluminum, which display resonant plasma oscillations in the visible range. As the plasma frequency is proportional to the square root of the carrier density, the electrodynamic equivalent of a metal at IR frequencies should have electron densities in the $10^{19}$ - $10^{20}$ cm$^{-3}$ range. To this aim we want to employ n-type germanium (n-Ge), grown on silicon wafers by Chemical Vapor Deposition (CVD). The use of n-Ge-on-Si is a breakthrough in IR plasmonics with respect to previous attempts based on doped silicon [2], since: (i) Ge-on-Si displays lower losses in the mid-IR and (ii) the small effective mass ($m^* = 0.12\ m_e$) of n-Ge provides higher plasma frequency for a given electron density.

## II. Material Growth

The samples were grown by low energy plasma enhanced chemical vapor deposition (LEPECVD). This growth technique is a variant of the conventional CVD characterized by the use of a low energy plasma to control the deposition of silicon-germanium alloys. Precursors gases (GeH$_4$ for germanium, PH$_3$ for doping) are introduced in the chamber, where the highly-reactive conditions created by the plasma cause the material to be efficiently deposited on the substrate. This gives great flexibility in the growth of high quality material, since the growth rate (controlled by the plasma density and by the amount of process gas) and the mobility of the adatoms (controlled by the substrate temperature) can be optimized separately. The sample was grown on a 100 mm p-Si(001) substrate with a resistivity of 5-10 Ω-cm. Before the heteroepitaxy, the native oxide was removed by dipping the substrate in aqueous hydrofluoric acid solution (HF :H$_2$O 1:10) for 30s. A 1 μm n-doped (~ $2.5 \times 10^{19}$ cm$^{-3}$) Ge layer was deposited at 500 °C at a growth rate of ~ 1 nm/s, with a GeH$_4$ flow of 20 sccm. The n-type doping was achieved in-situ by adding 0.15 sccm of PH$_3$.

## III. Infrared Spectroscopy

IR reflectance (*R*) measurements were performed on wafer portions with different doping levels at room temperature between 50 and 5000 cm$^{-1}$ with a FT-IR spectrometer (Bruker IFS66v). The sample was glued on a metal frame much larger than the spot size of the FT-IR (3 mm diameter). The frame was mounted on a vacuum manipulator inserted in the sample compartment of the FT-IR, and a home-made optical setup based on parabolic mirrors and broadband beamsplitters was used to shine light on the sample surface (n-Ge side) and to recollect the reflection from the same optical path, so to ensure perfectly normal incidence conditions. A co-aligned gold mirror was also mounted on the manipulator so to obtain the absolute reflectance of the n-Ge-on-Si multilayer, with no contributions from the sample holder in the frequency regions of non-zero transmission (NZT). The temperature of the sample was also varied down to T = 10 K, but the T-dependent data will be presented elsewhere.

Room-temperature results are presented in Fig.1b for two different samples. From now on, we will describe the spectrum of the sample with the highest doping level (yellow line, $n_e$ = 2.32 $10^{19}$ cm$^{-3}$ obtained from the IR data as explained below). Our discussion is however valid for all samples, provided that the frequency values are approximately rescaled by their $\omega_p$ ratio. For $\omega$ < 850 cm$^{-1}$ (region *A* of metal-like behavior) the reflectance *R* has a high absolute value. In this far-IR region, the skin depth is below 1 μm, hence much smaller than the n-Ge film thickness. Therefore, in this range the transmittance is zero, as the fraction (1-*R*) of the incoming radiation is entirely absorbed [3]. For 850 < $\omega$ < 1000 cm$^{-1}$ (region *B*) the reflectance shows a sharp drop which is indicative of the value of the screened plasma frequency $\omega_p = \sqrt{(4\pi n_e e^2/\varepsilon m^*)}$ = 1030 cm$^{-1}$, where $\varepsilon$ = 16 is the infinity dielectric constant of Ge. In this NZT region, the now larger fraction (1-*R*) of the impinging power is not entirely absorbed, and the transmittance of the multilayer is definitely not negligible, whereas the real part of the dielectric constant $\varepsilon_1(\omega)$ is still negative, as it will be shown at the end of this discussion by extracting the real and imaginary part of the dielectric constant self-consistently.

Oscillations in the reflectance above 1000 cm$^{-1}$ (region *C*) are due to multiple internal reflection interference within the thin Ge layer, schematically depicted in Fig. 1a, which has now become a dielectric-like material with $\varepsilon_1$ > 0. The fringe pattern is not a sinusoidal one, because the refractive index of n-Ge and, to a smaller extent, that of the lightly doped Si substrate are both frequency-dependent in the mid-IR.

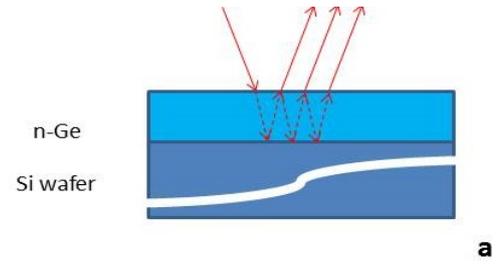

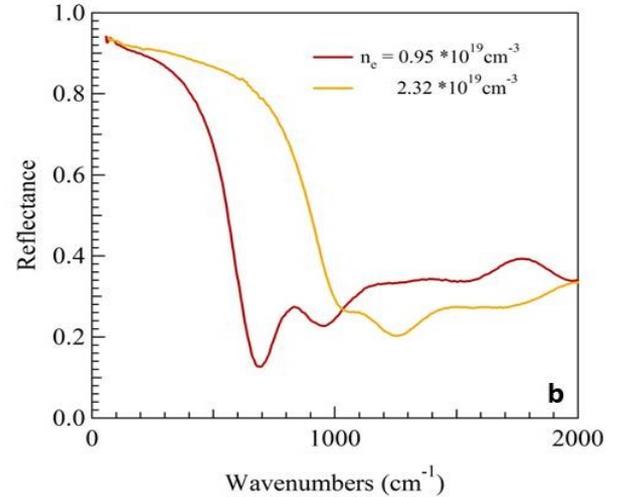

**Fig. 1. a** Sketch of the measured epitaxial layers: 2 μm thick nGe layer is grown over a Si wafer. **b** Reflectivity curves of two Ge-on-Si samples with different doping level as indicated in figure. Above the plasma frequency it is possible to see the interference fringes, compatible with the thickness of the Ge epilayer.

A multilayer fitting procedure, based on the Drude model, is used to reproduce the reflectance over the whole measured frequency range. This fitting procedure allows the retrieval of an analytical approximation for the dielectric constant of n-Ge which can be used to obtain reliable values of the activated carrier density $n_e$. On the other hand it is also known to be inadequate to describe the frequency-dependent electron-lattice scattering processes. Indeed the Drude model assumes a frequency-independent scattering rate which cannot describe electron scattering by phonons, nor by charged-impurities, that are the main sources of scattering in doped semiconductors. Noteworthy, the Drude-based fit can satisfactorily reproduce the data in regions A and C, but definitely not in region B. The reason for this is that the function $|\Delta R/\Delta \varepsilon|$ where $\Delta \varepsilon$ is the inaccuracy in reproducing the real part of the dielectric constant, diverges f or vanishing $\varepsilon$, *i.e.* in region *B*. Instead in regions A and C the model inaccuracy produces negligible effects on the reflectance if compared with the experimental error.

Once established that the dielectric constant of n-Ge can only be approximately derived from the reflectance data using the Drude model, we are now left with the problem of how to retrieve a model-independent dielectric constant of epitaxial n-Ge. The precise knowledge of the dielectric constant is of paramount importance for plasmonics applications. We can resort to the model-independent approach based on Kramers-Kronig (KK) transformations.

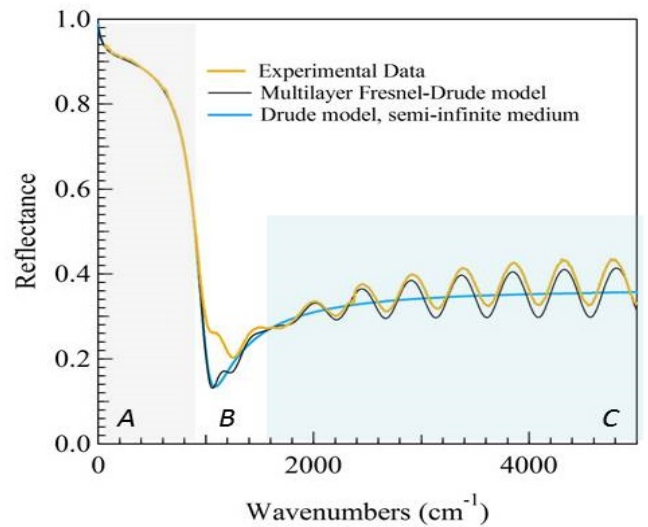

**Fig. 2.** Reflectivity curve of the highest doping Ge-on-Si sample compared with multilayer and infinite layer Drude fitting curve. In figure are reported the indications of frequency regions *A, B* and *C* described in the text.

KK transformations provide the full complex dielectric constant once the normal-incidence reflectance is known at all frequencies. We can then make a prolongation of the experimental data acquired in region A with the best-fit curve obtained with the Drude model, imposing an infinite thickness of the n-Ge layer (the data taken in region A on the 2 µm thick n-Ge film is not modified if the sample is actually infinitely thick, as can be observed from the comparison of the black and blue curves in Fig. 2). We then apply KK transformations to the interpolated prolonged data. From the obtained reflectance, $\bar{R}$, it is then possible to obtain the phase $\theta(\omega)$ and, subsequently, the real and imaginary parts of the dielectric constant $\tilde{\varepsilon}$ as:

$$\theta(\omega) = -\frac{2\omega}{\pi} P \int_{-\infty}^{+\infty} \frac{\ln\sqrt{\bar{R}}}{\omega'^2 - \omega^2} d\omega$$

$$\varepsilon_1(\omega) = \frac{\sqrt{1 - \bar{R} - 4 * \bar{R} * (\sin\theta)^2}}{1 + \bar{R} - 2\sqrt{\bar{R}} * (\cos\theta)^2}$$

$$\varepsilon_2(\omega) = \frac{4 * (1 - \bar{R})\sqrt{\bar{R}} * \sin\theta}{1 + \bar{R} - 2\sqrt{\bar{R}} * (\cos\theta)^2}$$

The obtained real and imaginary part of $\tilde{\varepsilon}$ are presented in Fig. 3 together with the energy loss function to highlight the position of the plasma frequency.

Due to the integral nature of the KK transformations, the uncertainty of the spectral shape of the dielectric constant at a given frequency is almost entirely related to the experimental uncertainty on $R(\omega)$ at that frequency. The uncertainty of the absolute value of the dielectric constant is due to the value of the $R(\omega)$ integrated over all frequencies. It can be demonstrated that the value of $\varepsilon_1$ and $\varepsilon_2$ in region A is accurate with a relative error of the order of $\Delta R/(1-R)$ where $\Delta R$ is the experimental uncertainty which is +/-2% in our experiment.

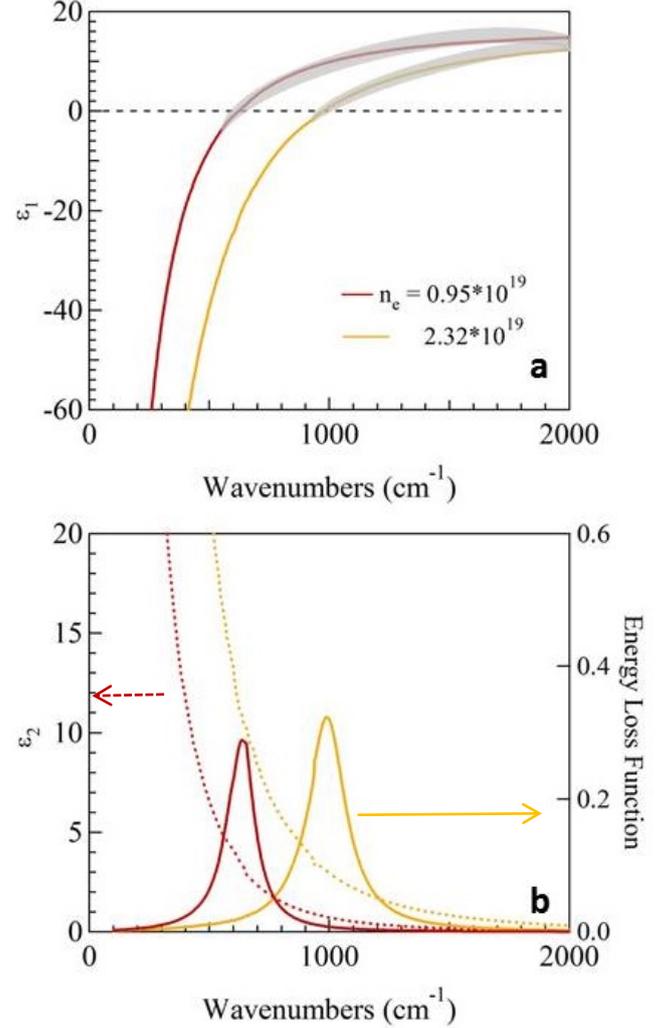

**Fig. 3** Real (a) and Imaginary (b) part of the dielectric constants as obtained by KK trasformations after data prolongation. The gray shaded area highlights the region corresponding to data prologation with the Drude model.

IV. SUMMARY

The electron-doped Ge-on-Si platform holds promise for developing cost-effective, CMOS-compatible plasmonics in the mid-IR. In order to exploit this material for proper electrodynamic design high precision dielectric constants are required. We have demonstrated a procedure to obtain the real and imaginary part of $\tilde{\varepsilon}$ with high accuracy by combining the standard multilayer fitting procedure based on the Drude model with Kramers-Kronig transformation of the absolute reflectance data in the zero-transmission range of the thin film.

V. ACKNOWLEDGMENTS

The research leading to these results has received funding from the European Union's Seventh Framework Programme under grant agreement n°613055.

REFERENCES

[1]. ] M. Ortolani, L. Baldassarre, A. Nucara, A. Samarelli, D.J. Paul, J. Frigerio, G. Isella, M. Finazzi and P. Biagioni, Infrared, Millimeter and THz waves, IEEE Xplore 10.1109/IRMMW-THz.2013.6665614 (2013)
[2] JC Ginn RL Jarecki Jr, EA Shaner, PS Davids, "Infrared plasmons on heavily-doped silicon" *J. Appl. Phys.* **110**, 043110 (2011).
[3]. M. Dressel and G. Gruener, "Electrodynamics of Solids" Cambridge University Press, 2002.